\documentclass[twocolumn,amsmath,amssymb,prl,aps,showpacs]{revtex4}
\usepackage{graphicx}
\usepackage{amsmath,amssymb}
\usepackage{mathrsfs}
\usepackage{color}
\usepackage{afterpage}
\usepackage{bm}
\usepackage[version=3]{mhchem}
\usepackage[normal]{subfigure}
\usepackage{enumitem}
\usepackage{hhline}

\newcommand{\R}{\mathbb{R}}

\begin{document}
\title{Big Data of Materials Science -- Critical Role of the Descriptor}
\author{Luca M. Ghiringhelli$^1$, Jan Vybiral$^2$, Sergey V. Levchenko$^1$, Claudia Draxl$^{3}$, and Matthias Scheffler$^1$}
\affiliation{$\phantom{ }^1$ Fritz-Haber-Institut der Max-Planck-Gesellschaft, Berlin-Dahlem, Germany \\
$\phantom{ }^2$ Charles University, Department of Mathematical Analysis, Prague, Czech Republic \\
$\phantom{ }^3$ Humboldt-Universit\"{a}t zu Berlin, Institut f\"{u}r Physik and IRIS Adlershof, Berlin, Germany
}

\date{\today}
\begin{abstract}
Statistical learning of materials  properties or functions so far starts with a largely silent, non-challenged step: the choice of the {\em set of descriptive parameters} (termed {\em descriptor}). However, when the scientific connection between the descriptor and the actuating mechanisms is unclear, causality of the learned descriptor-property relation is uncertain. Thus, trustful prediction of new promising materials, identification of anomalies, and scientific advancement are doubtful. We analyze this issue and define requirements for a suited descriptor. For a classical example, the energy difference of zincblende/wurtzite and rocksalt semiconductors, we demonstrate how a meaningful descriptor can be found systematically.
\end{abstract}
\pacs{02.60.Ed, 61.50.-f, 89.20.Ff, 31.15.E-}
\maketitle

Using first-principles electronic-structure codes, a large number of known and hypothetical materials has been studied in recent years, and currently, the amount of calculated data increases exponentially with time. Targets of these studies are, for example, the stable structure of solids or the efficiency of potential photovoltaic, thermoelectric, battery, or catalytic materials.
Utilizing such data like a reference book (query and read out what was stored) is an avail. Finding the actuating mechanisms of a certain property or function and describing it in terms of a {\em set of physically meaningful parameters} (henceforth termed {\em descriptor}) is the desired science.
A most impressive and influential example for the importance and impact of finding a descriptor is the periodic table of the elements, where the elements are categorized (described) by two numbers, the table row and column. The initial version had several ``white spots'', i.e., elements that had not been found at that time, but the chemical properties of these elements were roughly known already from their position in the table. Interestingly, the physical meaning of this two-dimensional descriptor became clear only later.
Below we will use an example from materials science to discuss and demonstrate the challenge of finding meaningful descriptors:
the prediction of the crystal structure of binary compound semiconductors, which are known to crystallize in zincblende (ZB), wurtzite (WZ), or rocksalt (RS) structures. The structures and energies of ZB and WZ are very close and for the sake of clarity we will not distinguish them here.
The energy difference between ZB and RS is larger, though still very small, namely just 0.001{\%} or less of the total energy of a single atom. Thus, high accuracy is required to predict this difference. 
This refers to both steps, the explicit calculations and the identification process of the appropriate descriptor (see below). The latter includes the representation of the descriptor-property relation.

For brevity, we only write ``property'', characterized by a number $\emph{{P}}_i$ in the following, with $i$ denoting the actually calculated material, but we mean the materials function(s) as well. In general, the property will be characterized by a string of numbers, but here we like to keep the discussion simple. Analogously, the multidimensional descriptor is denoted as a vector $\emph{\textbf{d}}_i$, with dimension $\Omega$.
The generalization of the discrete data set $\{P_i, \emph{\textbf{d}}_i\}$ to a continuous function $P(\emph{\textbf{d}})$ has been traditionally achieved in terms of physical models, or mathematical fits. Scientific understanding of the descriptor $\emph{\textbf{d}}$ and of the relationship between $\emph{\textbf{d}}$ and $\emph{{P}}$ is needed for deciding with confidence what new materials should be studied next as most promising novel candidates and for identifying interesting anomalies.

\begin{figure} [b]
\includegraphics[width=0.95\columnwidth,clip]{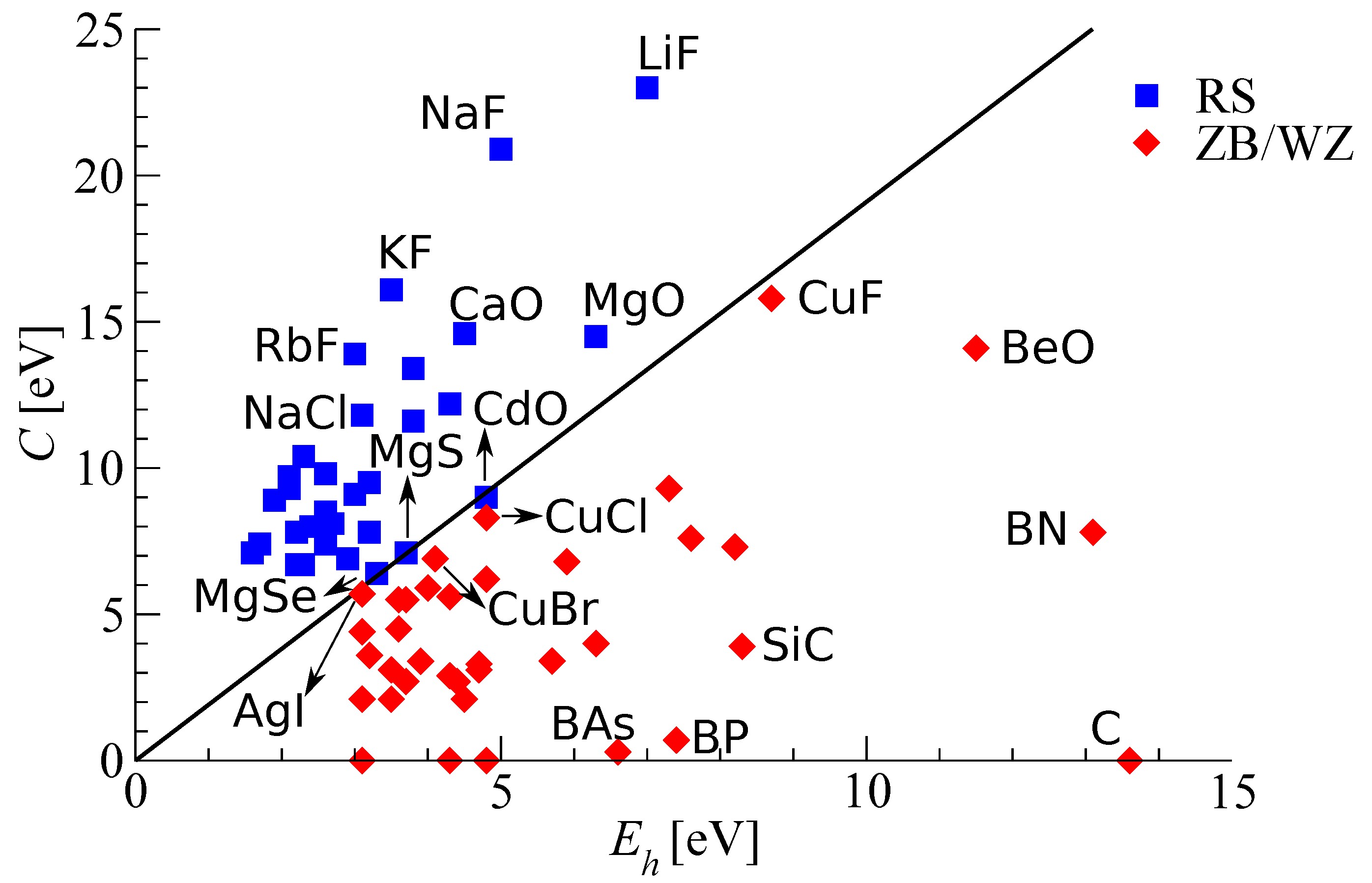}
\caption{Experimental ground-state structures of 68 octet binary compounds, arranged according to the two-dimensional descriptor introduced by Phillips and van Vechten \cite{vanVechten69,Phillips70}.
Because of visibility reasons, only 10 materials are labeled for each structure. See the SI for more details.}
\label{Fig:PvV}
\end{figure}

In 1970, Phillips and van Vechten (Ph-vV) \cite{vanVechten69,Phillips70} analyzed the 
classification challenge of ZB/WZ vs RS structures. They came up with a two-dimensional (2D) descriptor, i.e., two numbers that are related to the experimental dielectric constant and nearest-neighbor distance in the crystal \cite{vanVechten69,Phillips70}. Figure \ref{Fig:PvV} shows their conclusion. Clearly, in this representation ZB/WZ and RS structures separate nicely: Materials in the upper left part crystallize in the RS structure, those in the lower right part are ZB/WZ. Thus, based on the ingenious descriptor $\emph{\textbf{d}} = (E_h, C)$  one can predict the structure of unknown compounds without the need of performing explicit experiments or calculations. Several authors have taken up the Ph-vV challenge and identified alternative descriptors \cite{Zunger80,Pettifor84,Chelikowski12}. We will come back to this below.

In recent years, the demand for finding the function $P(\emph{\textbf{d}})$ employed statistical learning theory, which is the focus of this paper.
This strategy has been put forward by several authors in materials science \cite{Ceder06,Rajan08,Ramprasad13,Hart13,Mueller14}, as well as in bio- and cheminformatics (see, e.g., Ref. \cite{Rupp12} and references therein). Most of these works employed the kernel ridge regression (KRR) approach.
For a Gaussian kernel, the fitted property is expressed as a weighted sum of Gaussians: $P({\bm d}) = \sum_{i=1}^{N}  c_{i} \exp{ \left( - \| {\bm d}_{i} - {\bm d}\|^2_2 /  2\sigma^2 \right) } $, where $N$ is the number of training data points.
The coefficients $c_{i}$ are determined by minimizing  $ \sum_{i=1}^{N} (P({\bm d}_{i}) - P_{i})^2 + \lambda \sum_{i,j=1}^{N,N} c_{i}{c}_{j} \exp{ \left( - \| {\bm d}_{i} - {\bm d}_{j}\|^2_2 /  2\sigma^2 \right) } $, where  $\| {\bm d}_i - {\bm d}_j\|^2_2 = \sum_{\alpha=1}^\Omega ( d_{i,\alpha} - d_{j,\alpha})^2$ is the squared $\ell_2$ norm of the difference of descriptors of different materials, i.e., their ``similarity''.
The regularization parameter $\lambda$ and $\sigma$ are chosen separately, usually with the help of {\em leave-some-out cross validation} \cite{Tibshirani09}, i.e., by leaving some of the calculated materials out in the training process and testing how the predicted values for them agree with the actually calculated ones.

In essentially all previous materials studies the possibly multidimensional descriptor was introduced \emph{ad hoc}, i.e., without demonstrating that it was the best (in some sense) within a certain broad class (see Ref. \cite{Hart13} for an impressive exception).
In this Letter, we describe an approach for finding descriptors for the accurate prediction of a given property of a class of materials,
where we restrict ourselves to {\em ab initio} data.

For the example shown in Fig. \ref{Fig:PvV}, statistical learning is unnecessary, because one can determine the classification by visual inspection of the 2D plot. In this paper, we add the quantitative energy difference between ZB and RS to the original Ph-vV challenge. In general, the descriptor will be higher dimensional. Also the scientific question will be typically more complex than the structural classification. We define the conditions that a proper descriptor must fulfill in order to be suitable for causal ``learning'' of materials properties, and we demonstrate how the descriptor with the lowest possible dimensionality can be identified. Specifically, we will use the ``least absolute shrinkage and selection operator (LASSO)'' for {\em feature selection} \cite{LASSO}. 

All data shown in this study have been obtained with density-functional theory using the local-density approximation (LDA) for the exchange-correlation interaction. Calculations were performed using the all-electron full-potential code FHI-aims \cite{Blum09} with highly accurate basis sets, $\bf k$-meshes, and integration grids.
For the task discussed in this paper, the quality of the exchange-correlation functional is irrelevant. Nevertheless, we stress that the LDA provides a good description of the studied materials.  In particular, we have computed  equilibrium lattice constants and total energies for all three considered lattices (ZB, WZ, RS) of a set of 82 binary materials. The full list of these materials and all calculated properties can be found in the SI and all in- and output files can be downloaded from the NoMaD repository \cite{comment_nomad}. Furthermore, we calculated several properties of the isolated neutral atoms and dimer molecules (see below).

\begin{figure} [h!]
\centering
\includegraphics[width=0.97\columnwidth,clip]{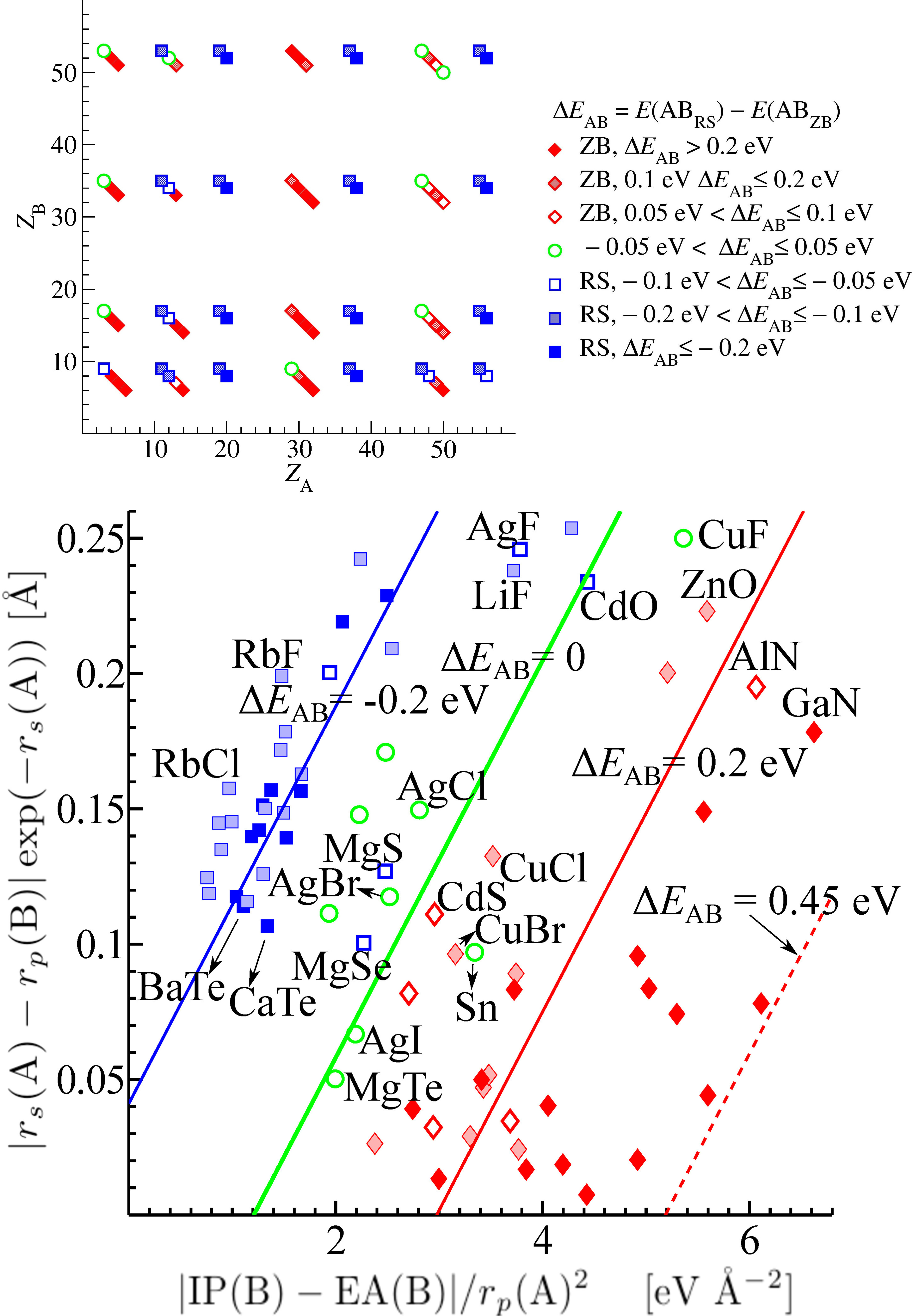}
\caption{Calculated energy differences between RS and ZB structures of the 82 octet binary AB materials, arranged by using the nuclear numbers $(Z_\textrm{A}, Z_\textrm{B})$ as descriptor (top) and 
according to our optimal two-dimensional descriptor (bottom). 
In the bottom panel, seven ZB materials with predicted $\Delta E_\textrm{AB} > 0.5$~eV are outside the shown window (see SI). }
\label{Fig:JV}
\end{figure}

Let us start with a simple example that demonstrates
the necessity of {\em \textbf{validation}} in the search for descriptors.
The nuclear numbers of a binary semiconductor AB, $Z_\textrm{A}$ and $Z_\textrm{B}$, unambiguously identify the lowest energy structure: They define the many-body Hamiltonian, and its total energies for the different structures give the stable and metastable structures. Figure \ref{Fig:JV} (top) displays the total-energy differences of the ZB and RS structures as function of $Z_\textrm{A}$ and $Z_\textrm{B}$. When using the KRR approach the data can be fitted well (see SI) when the whole set is used for learning. However, the predictive power of KRR based on the descriptor $\emph{\textbf{d}} = (Z_\textrm{A}, Z_\textrm{B})$ is bad, as tested by leave-some-out cross validation (see Table I and SI).
Obviously, the relation between $\emph{\textbf{d}} = (Z_\textrm{A}, Z_\textrm{B})$ and the property that we need to learn is by far too complex.

For a descriptor, we consider the following properties to be important, if not necessary: 
\renewcommand{\labelenumi}{\alph{enumi})}
\begin{enumerate}[noitemsep,nolistsep]
\item A descriptor {\em \textbf{d}}$_i$
uniquely characterizes the material $i$ as well as property-relevant elementary processes.
\item  Materials that are very different (similar) should be characterized by very different (similar) descriptor values.
\item The determination of the descriptor must not involve calculations as intensive as those needed for the evaluation of the property to be predicted.
\item The dimension $\Omega$ of the descriptor should be as low as possible (for a certain accuracy request). 
\end{enumerate}
Although the  Ph-vV descriptor $\emph{\textbf{d}} = (E_h, C)$ fulfills conditions a), b), and d), it falls short on condition c). In contrast, $\emph{\textbf{d}} = (Z_\textrm{A}, Z_\textrm{B})$ fails for conditions b) and d).

In order to identify a good descriptor, we start with a large number $M$ of candidates (the ``feature space'') for the components of ${\bm d}$. We then look for the $\Omega$-dimensional ($\Omega=1,2,\ldots$) descriptor $\emph{\textbf{d}}$ that gives the best linear fit of
$P(\emph{\textbf{d}})$: $ P(\emph{\textbf{d}}) = {\bm d}{\bm c} $, where ${\bm c}$ is the $\Omega$-dimensional vector of coefficients. It is determined by minimizing the {\em loss} function $\| \emph{\textbf{P}} -  \emph{\textbf{D}} \emph{\textbf{c}} \|_2^2$, where ${\bm D}$ is a matrix with each of the $N$ rows being the descriptor ${\bm d}_i$ for each training data point, and ${\bm P}$ is the vector of the training values $P_i$. 
We emphasize that the choice of a linear fitting function for $P(\emph{\textbf{d}})$ is not restrictive since, as we will show below, non-linearities are included in a controlled way in the formation of the candidate components of ${\bm d}$.
The function $P({\bm d})$ is then determined by only $\Omega$ parameters.

The task is now to find, among all the $\Omega$-tuples of candidate features, the $\Omega$-tuple that yields the smallest $\| \emph{\textbf{P}} -  \emph{\textbf{D}} \emph{\textbf{c}} \|_2^2$.
Unfortunately, a computational solution for such problem is infeasible (NP-hard) \cite{Arora09}.
LASSO \cite{LASSO} provides sparse (i.e., low-dimensional) solutions by recasting the NP-hard problem into a convex minimization problem
\begin{equation}\label{eq:rr1}
\mathop{\rm arg min}\limits_{ \emph{\textbf{c}} \in \R^{M}} \|  \emph{\textbf{P}} -  \emph{\textbf{D}} \emph{\textbf{c}} \|_2^2+\lambda\|  \emph{\textbf{c}}\|_1, 
\end{equation}
where the use of the $\ell_1$-norm ($\| \emph{\textbf{c}} \|_1 =\sum_{\alpha=1}^M|c_\alpha|$) is crucial.
The larger we choose $\lambda>0$, the smaller is the $\ell_1$-norm of the solution of Eq. \ref{eq:rr1} and vice versa.
There is actually a smallest $\bar{\lambda}>0$, such that the solution of Eq. \ref{eq:rr1} is zero. If $\lambda<\bar{\lambda}$, one or more coordinates of $ \emph{\textbf{c}}$ become non-zero.

\begin{table}[th!]
\centering
\begin{tabular}{l||c|c||c|c|c|c|c|}
 Descriptor & $Z_\textrm{A}, Z_\textrm{B}$ & $r_\sigma, r_\pi$ & 1D & 2D & 3D & 5D \\
\hline
RMSE & $2\! \cdot \!10^{-4}$ & 0.07 & 0.14 & 0.10 & 0.08 & 0.06 \\
MaxAE & $8\! \cdot \!10^{-4}$ & 0.25 & 0.32 & 0.32 & 0.24 & 0.20 \\
\hline
RMSE, CV & 0.19 & 0.09 & 0.14 & 0.11 & 0.08 & 0.07 \\
MaxAE, CV & 0.43 & 0.17 & 0.27 & 0.18 & 0.16 & 0.12 \\
\hline
\end{tabular}
\caption{Root mean square error (RMSE) and maximum absolute error (MaxAE) in eV for the least-square fit of all data (first two lines) and for the test set in a leave-10\%-out cross validation (L-10\%-OCV), averaged over 150 random selections of the training set (last two lines). The errors for $(Z_\textrm{A}, Z_\textrm{B})$ and $(r_\sigma, r_\pi)$ \cite{Zunger80} are for a KRR fit at hyperparameters $(\lambda,\sigma)$ that minimize the RMSE for the L-10\%-OCV (see SI). The errors for the $\Omega=1, 2, 3, 5$ (noted as 1D, 2D, 3D, 5D) descriptors are for the LASSO fit. In the L-10\%-OCV for the latter descriptors, the overall LASSO-based selection procedure of the descriptor (see text) is repeated at each random selection of the test set.}
\end{table}

We note that the so-called ``feature selection'' is a widespread set of techniques that are used in statistical analysis in different fields \cite{FitSel}, and LASSO is one of them. LASSO was successfully demonstrated in Ref. \cite{Hart13}, for identifying the low-dimensional representation of the formation energy of an alloy, within the cluster expansion of the Hamiltonian. Obviously, when a well identified basis set, such as the cluster expansion, is not available for the property to be modeled, the feature space must be constructed differently. In this paper, we start from scientific insight, i.e., defining physically motivated {\em primary features} that form the basis for a large feature space. We then search for a low-dimensional descriptor that minimizes the RMSE, given by $\sqrt{(1/N) \| \emph{\textbf{P}} -  \emph{\textbf{D}} \emph{\textbf{c}} \|_2^2}$, for our $N\!=\!82$ binary compounds. The property $P$ that we aim to predict is the difference in the LDA energies between RS and ZB for the given atom pair AB, $\Delta E_\textrm{AB}$. The order of the two atoms is such that element A has the smallest Mulliken electronegativity: $\textrm{EN}\! =\! -(\textrm{IP}+\textrm{EA})/2$. IP and EA are atomic ionization potential and electron affinity. 

For constructing the feature space, i.e., the candidate components of the descriptor, and then selecting the most relevant of them, we implemented an \textbf{\em iterative} approach. At first we defined {\em primary features}. These are (for atom A): IP(A) and EA(A), H(A) and L(A), i.e., the energies of the highest-occupied and lowest-unoccupied Kohn-Sham (KS) levels, as well as $r_s$(A), $r_p$(A), and $r_d$(A), i.e., the radii where the radial probability density of the valence $s$, $p$, and $d$ orbitals are maximal. The same was done for atom B. In addition to these atomic data, we offered information on AA, BB, and AB dimers, namely their equilibrium distance, binding energy, and HOMO-LUMO KS gap. Altogether, these are 23 {\em primary features}.

Next, we define rules for linear and non-linear combinations of the {\em primary features}. One can easily generate a huge number of candidate descriptors, e.g., all thinkable but not violating basic physical rules. In the present study,
we used about $10\,000$ candidates grouped in subsets that are used in the different iterations (see SI). A more detailed discussion will be given in Ref. \cite{next}.
In the language of KRR, this approach designs a kernel, here done by using physical insight. 
Not surprisingly, LASSO (and actually any other method) has difficulties in selecting among highly correlated features \cite{comment4}.
In these cases, it is not ensured that the first $\Omega$ selected features form the best $\Omega$-dimensional descriptor.
Although checking correlations between pairs is straightforward and computationally reasonably inexpensive, discovering correlations between triples and more-tuples is computationally prohibitive.
Therefore, we adopted a different strategy: The first 25-30 features proposed by LASSO were selected and a batch of least-square fits performed (when the descriptor is fixed, i.e. the non-zero components of $\emph{\textbf{c}}$ are fixed, Eq. \ref{eq:rr1} reduces to a linear, least-square, fit), taking in turn as ${\bm D}$ each single feature, each pair, etc. We confirmed that this strategy always finds the best descriptor by running the mentioned extensive search for several different subsets of hundreds of features.

Our procedure identifies as best (i.e., lowest-RMSE) 1D, 2D, and 3D descriptors, the first, the first two, and all three of the following features:
\begin{equation}
  \frac{\textrm{IP(B)}-\textrm{EA(B)}}{r_p(\textrm{A})^2},\, \frac{|r_s(\textrm{A})-r_p(\textrm{B})|}{\exp (r_s(\textrm{A}))},\, \frac{\left| r_p\textrm{(B)}-r_s\textrm{(B)} \right|}{\exp(r_d\textrm{(A)})}.
\label{Eq:desc}
\end{equation}

\noindent Note that, mathematically, the descriptor does not necessarily need to build up incrementally in this way, e.g., the 1D one may not be a component of the 2D one. However, in our study, it does. 
The RMSE and MaxAE for the 1D, 2D, 3D descriptors are reported in Table I. By adding further dimensions to the descriptor, the decrease of the RMSE becomes smaller and smaller.

We tested the robustness of our descriptor by performing a leave-10\%-out cross validation (L-10\%-OCV). Thereby, the overall procedure of selecting the descriptor is repeated from scratch on a learning set obtained by randomly selecting 90\% of the materials. The resulting fitted linear model is applied to the excluded materials and the prediction errors on this set, averaged over 150 random selections, are recorded. The results are shown in Table I. Not only the RMSE, but also the selection of the descriptor proved very stable. In fact, the 2D descriptor was selected 100\% of the times, while the 1D descriptor was the same in 90\% of the cases.

The errors for the 2D descriptor introduced by Zunger (Refs. \cite{Zunger80,Chelikowski12} and SI), based on sums and absolute differences of $r_s$'s and $r_p$'s, are also reported in Table I.
The cross-validation error of the linear fit with our 2D descriptor is as small as the highly non-linear KRR-fit with Zunger's 2D descriptor.
However, our descriptor bears the advantage that it was derived from a broad class of options by a well-defined procedure, providing a basis for a systematic improvement (with increasing $\Omega$).
Our LASSO-derived descriptor contains physically meaningful quantities, like the band gap of B in the numerator of the first component and the size mismatch between valence $s$- and $p$-orbitals (numerators of the second and third component). We note that the components of the descriptors are not symmetric wrt exchange between A and B. Symmetric features were included in the feature space, but never emerged as prominent and, for the selected descriptors, symmetrized versions were explicitly constructed, tested, and systematically found performing worse. This reflects that the symmetry was explicitly broken in the construction of the test set, as the order AB in the compound is such that $\textrm{EN(A)} < \textrm{EN(B)}$. Furthermore, we find that $d$ orbitals appear only in the third or higher dimension.
In Fig. \ref{Fig:JV} (bottom)  we show the calculated and predicted $\Delta E_\textrm{AB}$, according to our best 2D descriptor.
It is evident that our 2D descriptor fulfills all above noted conditions, where conditions a), c), and d) are in fact ensured by construction.

In order to further test the robustness and the physical meaningfulness of the identified descriptor, we performed tests by perturbing the value of the property $\Delta E_\textrm{AB}$ by adding uniform noise in the interval $[-0.1, 0.1]$ eV. The 2D descriptor of Eq. \ref{Eq:desc} was identified 93\% of the times, with an increase of the RMSE by 10\% only. More details are reported in Ref. \cite{next}.
This test shows that the model allows for some uncertainty in the measure property.
Larger noise terms, however, destroy the reliable identification of the descriptor (see Ref. \cite{next}).
This analysis implies that the descriptor identified by LASSO
contains the important physically meaningful ingredients for the prediction of $\Delta E_\textrm{AB}$, even though a physical model that justifies the $P{(\bm d})$ mapping is not transparent.

We finally comment on the causality of the learned descriptor-property relationship. 
The simplicity of our model is in sharp contrast with what is yielded by, for instance, KRR, where as many fit parameters as observed points are, in principle, necessary. 
As an indication of having identified a causal (physically meaningful) descriptor for the property $\Delta E_\textrm{AB}$, we use the stability of the selection of the descriptor upon both L-10\%-OCV and perturbation of the values of the property, under the condition that the $P{(\bm d})$ dependence has a small number of fit parameters and a simple functional form (see Eq. \ref{Eq:desc} and SI). 

Financial support from the Einstein Foundation Berlin is appreciated. JV acknowledges financial support of the ERC CZ grant LL1203.
We thank Krishna Rajan for bringing the relevance of the Ph-vV analysis to our attention, and Judea Pearl for inspiring discussions on the concept of causality in the context of statistical inference.


\begin{thebibliography}{19}
\bibitem{comment4}
Two columns of ${\bm D}$ are correlated if the absolute value of their Pearson's correlation index is (about) 1.
\end{thebibliography}
\end{document}